\documentclass[10pt,conference,letterpaper]{IEEEtran}
\IEEEoverridecommandlockouts
\usepackage{cite}

\usepackage{amsmath,amssymb,amsfonts}
\usepackage{algorithmic}
\usepackage{graphicx}
\usepackage{textcomp}
\usepackage{xcolor}
\usepackage[ruled,linesnumbered]{algorithm2e} 
\usepackage[nolist]{acronym}
\usepackage{soul}
\usepackage{marginnote}
\usepackage{url}
\usepackage[inline]{enumitem}
\usepackage{tabularx}
\usepackage{amsthm}
\usepackage{tikz}
\usepackage[utf8]{inputenc}
\usepackage[T1]{fontenc}

\newcommand{\myssec}[2]{%
  \subsection{#1}%
  \label{sec:#2}%
}
\newcommand{\rsec}[1]{%
  Sec.~\ref{sec:#1}%
}

\newcommand{\removed}[1]{%
  {\color{red}#1%
  }%
}
\renewcommand{\removed}[1]{}

\newcommand*{\thead}[1]{\multicolumn{1}{c}{\bfseries #1}}
\newcommand{\myfigeps}[3][width=\columnwidth]{%
\begin{figure}[tbp]%
\centering%
\includegraphics[#1]{figures/#2}%
\vspace{-1em}%
\caption{#3}%
\vspace{-1em}%
\label{fig:#2}%
\end{figure}%
}
\newcommand{\myfigdoubleeps}[5][width=\columnwidth]{%
\begin{figure*}[tb]%
\centering%
\includegraphics[#1]{figures/#2}%
\includegraphics[#1]{figures/#3}%
\vspace{-1em}%
\caption{#5}%
\vspace{-1em}%
\label{fig:#4}%
\end{figure*}%
}
\newcommand{\myfigfulleps}[3][width=\textwidth]{%
\begin{figure*}[tb]%
\centering%
\includegraphics[#1]{figures/#2}%
\caption{#3}%
\label{fig:#2}%
\end{figure*}%
}

\newcommand{\rfig}[1]{Fig.~\ref{fig:#1}}
\newcommand{\ralgo}[1]{Algorithm~\ref{algo:#1}}

\newcommand{\rtab}[1]{Table~\ref{tab:#1}}
\newcommand{\req}[1]{Eq.~(\ref{eq:#1})}

\newenvironment{myinlinelist}%
{%
\begin{enumerate*}[label=(\roman*)]%
}%
{%
\end{enumerate*}%
}

\newenvironment{myitemlist}%
{%
\begin{itemize}[parsep=0em,leftmargin=*,label={--}]%
}%
{%
\end{itemize}%
}

{%
\begin{enumerate}[parsep=0em,leftmargin=*,label=\arabic*.]%
}%
{%
\end{enumerate}%
}

\begin{document}

\title{
Resource Allocation in Quantum Networks\\for Distributed Quantum Computing
}

\author{
\IEEEauthorblockN{Claudio Cicconetti}
\IEEEauthorblockA{\textit{IIT, CNR} --
Pisa, Italy \\
c.cicconetti@iit.cnr.it}
\and
\IEEEauthorblockN{Marco Conti}
\IEEEauthorblockA{\textit{IIT, CNR} --
Pisa, Italy \\
m.conti@iit.cnr.it}
\and
\IEEEauthorblockN{Andrea Passarella}
\IEEEauthorblockA{\textit{IIT, CNR} --
Pisa, Italy \\
a.passarella@iit.cnr.it}
}

\author{Claudio~Cicconetti,
        Marco~Conti,
        and Andrea~Passarella%
\IEEEcompsocitemizethanks{\IEEEcompsocthanksitem All the authors are with the Institute of Informatics and Telematics (IIT) of the National Research Council (CNR), Pisa, Italy.}%
}

\IEEEtitleabstractindextext{%
\begin{abstract}
  The evolution of quantum computing technologies has been advancing at a steady
pace in the recent years, and the current trend suggests that it will
become available at scale for commercial purposes in the near future.
The acceleration can be boosted by pooling compute infrastructures
to either parallelize algorithm execution or solve bigger instances
that are not feasible on a single quantum computer, which requires
an underlying \textit{Quantum Internet}: the
interconnection of quantum computers by quantum links and repeaters
to exchange entangled quantum bits.
However, Quantum Internet research so far has been focused on provisioning
point-to-point flows only, which is suitable for (e.g.) quantum sensing and
metrology, but not for distributed quantum computing.
In this paper, after a primer on quantum computing and networking,
we investigate the requirements and objectives of smart computing
on distributed nodes from the perspective of quantum network
provisioning.
We then design a resource allocation strategy that is evaluated
through a comprehensive simulation campaign, whose results highlight
the key features and performance issues, and lead the way to further
investigation in this direction.

\end{abstract}

}

\maketitle

\begin{tikzpicture}[remember picture,overlay]
\node[anchor=south,yshift=10pt] at (current page.south) {\fbox{\parbox{\dimexpr\textwidth-\fboxsep-\fboxrule\relax}{
  \footnotesize{
    \textcopyright 2022 IEEE.  Personal use of this material is permitted.  Permission from IEEE must be obtained for all other uses, in any current or future media, including reprinting/republishing this material for advertising or promotional purposes, creating new collective works, for resale or redistribution to servers or lists, or reuse of any copyrighted component of this work in other works.
  }
}}};
\end{tikzpicture}

\IEEEdisplaynontitleabstractindextext

\IEEEpeerreviewmaketitle

\bibliographystyle{abbrv}


%
  \section{Introduction}%
  \label{sec:introduction}%
  \ac{QC} exploits the
properties of matter at very small scale to solve some problems
much faster than a classical counterpart.
Even though \ac{QC} has been theorized 40 years
ago~\cite{preskill_quantum_2021}, only recently the technology
evolution and a spur of investments have made it possible to obtain
practical results and speculate about approaching mass
deployments~\cite{sevilla_forecasting_2020}.
\ac{QC} is being already used in the chemical and pharmaceutical
industry, while new applications are being progressively unlocked in
material science, \ac{ML} and engineering optimization, production
and logistics, post-quantum security
\cite{quantum_technology_and_application_consortium__qutac_industry_2021}.
Essentially, the computational advantage of \ac{QC} stems from the
properties of superposition and entanglement of the \textit{qubits}
(i.e., the ``quantum bits''), which we review in \rsec{basics}.
%
%
%

We can expect that the computational power of a single \ac{QC} will
remain relatively limited in the near future, due to scalability
issues in maintaining a very stable and controlled environment to
cope with the flimsy nature of qubits.
For this reason, computation speed-up can be sought by distributing
the execution over multiple \acp{QC}, which requires an end-to-end
entanglement of the qubits they use across geographical distances,
in other terms the realization of the \textbf{Quantum
Internet}~\cite{wehner_quantum_2018}.
The latter, in fact, is receiving
attention and investments: when deployed, it will allow \acp{QC}
to run algorithms in a distributed fashion, thus benefiting the
users from shared compute capacity across multiple systems, much
like what happened with grid computing in the not-so-distant past.
%
%

In this work we focus on a problem that is ancillary to distributed
\ac{QC} and has received little attention so far: the allocation
of resources in the Quantum Internet among multiple quantum computers based
on the characteristics of the underlying quantum network.
%
%
The properties of distributed \ac{QC} applications are fundamentally
different from those of point-to-point quantum applications, which have
been studied intensively by the scientific community in relation to
quantum sensing~\cite{pirandola_advances_2018} and
\ac{QKD}~\cite{pedone_toward_2021}.
In particular, distributed \ac{QC} applications have a more elastic
nature along two directions:
\begin{myinlinelist}
    \item a given host may have multiple communication peers, as a
    quantum computer may offload its computation to many nodes
    in a pool, not just one; and
    \item the rate of entangled qubits exchanged is not constant,
    but rather a quantum computer may desire to consume as many of
    them as possible in a greedy manner, which leads to fairness
    concerns.
\end{myinlinelist}
\textit{
Our main contribution is to explore these novel requirements from
the point of view of network resource allocation,
while also proposing a practical solution, which we evaluate through
simulations with the goal of identifying performance trade-offs in
terms of internal and external system properties.
}

The rest of this paper is structured as follows.
In \rsec{basics} we provide an introduction to
quantum computing and networking.
We review the related work on routing in the Quantum Internet in
\rsec{soa}.
We then describe the system model adopted and our scientific
contribution in \rsec{contribution}.
Simulation results are presented in \rsec{eval}, while
\rsec{conclusions} concludes the paper.
%

%
  \section{Basics of Quantum Computing and Communication}%
  \label{sec:basics}%
  In this section we provide a bird's-eye view of the fundamental
principles of quantum computing and communication, with the goal
of making the paper readable even without prior knowledge on these
topics.
We suggest the following textbooks to the interested reader for a complete
illustration of quantum computing~\cite{nielsen_quantum_2010} and
networking~\cite{van_meter_quantum_2014}.

As already introduced briefly, the unit of computation of \ac{QC} is
the \textit{qubit}.
While a classical bit represents a binary piece of information, e.g.,
\texttt{head} (0) or \texttt{tail} (1) as the result of flipping a coin,
a qubit can be in a \textit{superposition} of the two opposite states,
which can be visualized as the coin flipping in the air: until it
drops, it is impossible to say what is its value.
In conventional bra-ket notation this is expressed as
$|\psi\rangle = \alpha |0\rangle + \beta |1\rangle$, where $|0\rangle$ and
$|1\rangle$ are the two extreme levels of the systems (\texttt{head} or
\texttt{tail} in the example) and $\alpha,\beta$ are complex numbers
representing their respective probability amplitudes.
Measuring the qubit causes its state to collapse to one of the two
possible values, thus effectively converting the qubit into a classical bit;
once a qubit has been measured its quantum information cannot be recovered
anymore (i.e., once the coin has dropped, it is not possible to restore it
to its status while flipping).
Quantum algorithms are executed by preparing the input qubits in known
initial states, based on the configuration of the real-life problem that one is
going to solve, then applying quantum gates expressing quantum logic
operations, and finally measuring the outcomes, which gives the output as
classical information.
%

Another important property of qubits is that they can be \textit{entangled}.
When a set of qubits is entangled, they express a high correlation that has
no classical counterpart (and, in fact, is deeply counterintuitive), which
persists even though they are separated in space.
%
For instance, let us consider two qubits in the maximally entangled
\textit{Bell state}:
\begin{equation}\label{eq:bell-state}
    |\Phi^+\rangle := \frac{1}{\sqrt{2}}\left(|00\rangle + |11\rangle\right).
\end{equation}
When performing a measurement on any of the two qubits, there is
equal probability to get $0$ or $1$, but if the measurements of the
first qubit gives (e.g.) $0$, then measuring the second qubit will
certainly given a $0$; this remains true even if the
second measurement is done later or at an arbitrary distance.

\myfigeps{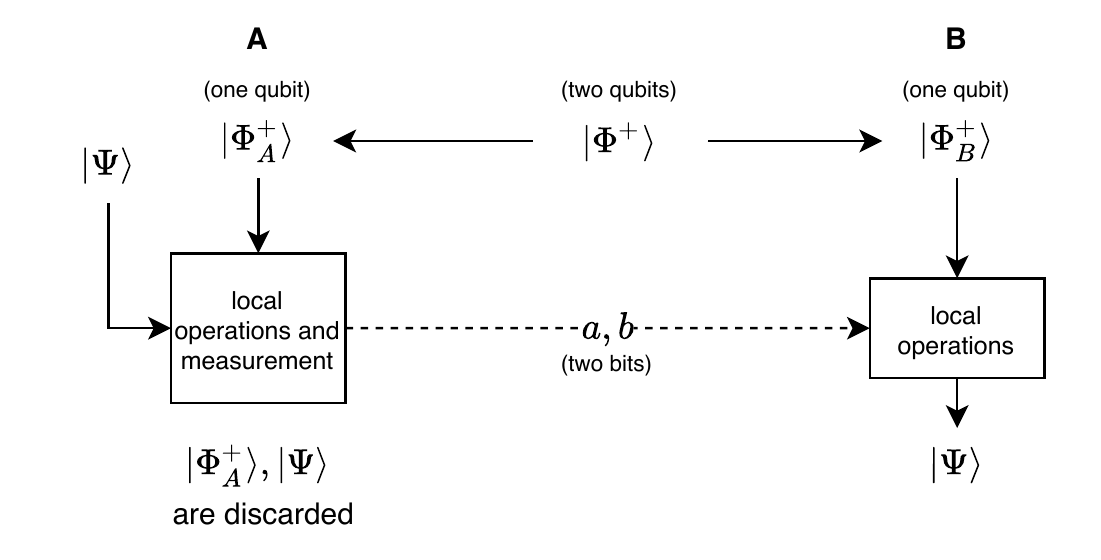}{Quantum teleportation scheme.
$|\Psi\rangle$ is the qubit to be teleported from A to B, while
$|\Phi_A^+\rangle$ and $|\Phi_B^+\rangle$ are two maximally entangled
qubits in a Bell state that are consumed in the process; $(a,b)$ are
classical bits that are read by A and used by B to decide which local
operations to apply.}

Another distinguishing feature of qubits is that they \textit{cannot} be
copied:
%
a qubit must be either transferred \textit{as is} or ``consumed'' through
(irreversible) measurement.
However, it is possible to \textit{teleport} a qubit from a quantum system (A)
to another (B) as follows -- see scheme in \rfig{basics-1}.
The two quantum systems must have
exchanged an entangled pair of qubits in the Bell state beforehand
(call them $|\Phi_A^+\rangle$ and $|\Phi_B^+\rangle$); then, A performs local
operations on $|\Phi_A^+\rangle$ and the qubit to teleport $|\Psi\rangle$,
followed by a measurement, which destroys both.
Such a measurement provides a result, expressed
as two classical bits ($a,b$), which are transferred to B.
Finally, B applies local operations
on $|\Phi_B^+\rangle$, which transform it exactly into $|\Psi\rangle$.
It is important to note that the duration of teleporting is limited by the
time required to transfer classical information from A to B, hence it does
not violate the laws of physics that prohibit faster-than-light travel
(even for information).
%


Teleportation is a crucial operation involved in the creation of
quantum networks.
%
%
Quantum computers store so-called \textit{matter qubits}
in quantum memories, which are interconnected by logic that makes
them interact via quantum gates for the execution of algorithms.
%
%
To extend the execution of a quantum circuit beyond a single computer,
it is necessary to create an entanglement between one of its qubits with
another, physically located in the quantum memory of a remote peer.
This can be achieved through so-called \textit{flying qubits}, typically
realized by encoding the state into the polarization of
photons, which can be transmitted efficiently over fiber optic
cables~\cite{tomm_bright_2021} or in free-space using satellite
links~\cite{khatri_spooky_2021}.

\myfigfulleps{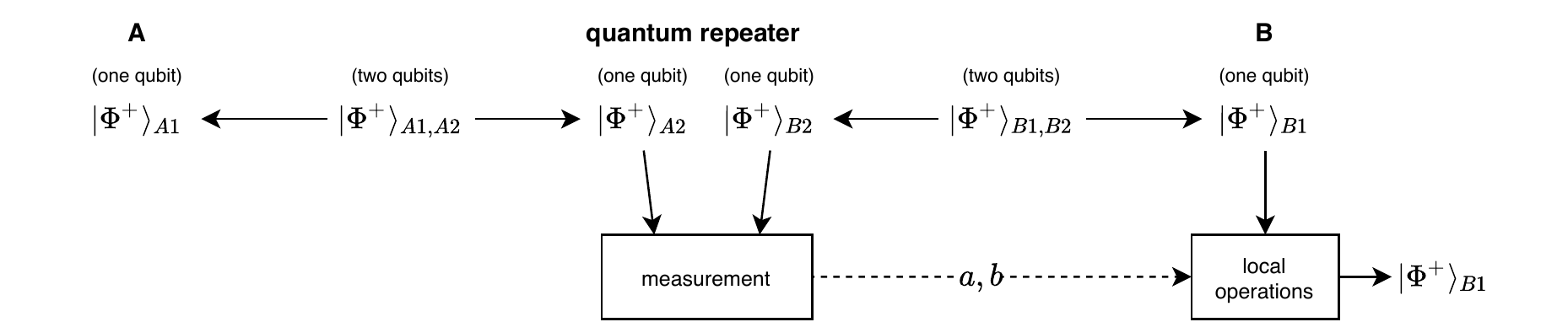}{Entanglement swapping scheme.
A pair of maximally entangled qubits $|\Phi^+\rangle_{A1}$,
$|\Phi^+\rangle_{B1}$ is generated thanks to the measurement
operation done by an intermediate quantum repeater, followed by
local operations on, e.g., B.}

However, flying qubits fade with distance, which makes it impractical
to deploy a global scale quantum network
without intermediate devices that can extend the range of single links:
the \textbf{quantum repeaters}.
A quantum repeater is a device that performs a measurement, called
\textit{entanglement swapping}, between two
(flying) qubits resulting in the end-to-end entanglement of two (matter) qubits,
as illustrated in \rfig{basics-3}, following the same principle of
teleportation discussed above.
%
%
The entanglement swapping can be repeated
along a chain of quantum repeaters between two end points, thus
extending the range of quantum networks for the creation of
end-to-end entangled pairs of qubits arbitrarily, at least in theory.
In practice, channel impairments and measurement imperfections
reduce the quality of the end-to-end entanglement generated.
A widely used metric to measure these collective effects is the
\textit{fidelity}, which is a relative measure between the actual
state of a quantum system and its desired state; the fidelity is
defined in the range $[0,1]$, where $1$ means perfection.
Practical quantum algorithms can tolerate some degree of fidelity degradation.
%
A comprehensive review of quantum teleportation for the realization of
the Quantum Internet can be found in~\cite{cacciapuoti_when_2020}.

The evolution of quantum networks is commonly seen in incremental
steps, sometimes referred as ``generations'' (term inspired from
cellular networks), where the first generation (1G) is characterized
by quantum repeaters operating according to the scheme above.
Future generations will embed error correction to ensure high-fidelity
entangled pairs at long distance and with many intermediate
nodes~\cite{muralidharan_optimal_2016}, but we do not consider them
as they are very far from being available in practice ---indeed,
not even 1G quantum repeaters are available today as commercial
off-the-shelf products.
%

Finally, the attentive reader might have noticed
that entanglement swapping does not
lead to the distribution of arbitrary quantum states $|\psi\rangle$,
but only Bell states as in \req{bell-state}.
Such an assumption does \textit{not} restrict the general applicability of a
network of quantum repeaters to realize distributed \ac{QC}, because it
can be shown that \textit{any arbitrary state can be obtained from a Bell state
through a sequence of local operations only}.
  \section{Related Work}%
  \label{sec:soa}%
  The literature on quantum networking and distributed \ac{QC} is not vast:
even though the basic ingredients
have been known since a long time ago ---consider for instance the
seminal paper by Bouwmeester~\textit{et al.} on quantum
teleportation~\cite{bouwmeester_experimental_1997} published on Nature
in 1997--- only recently there have been investments in an order of
magnitude sufficient for technology to take off.
This revamped interest has triggered new research
activities in this area, briefly reviewed below.

In general terms, the problem of \textbf{quantum routing} is formulated
as follows: given a network of quantum nodes (repeaters or computers) and
a set of traffic flows identified by their sources, destinations, and
application requirements (e.g., the minimum fidelity),
find the ``best'' paths that fulfill the constraints.
Some works have studied the problem by reusing the findings in the
area of routing in classical networks.
Van Meter~\textit{et al.} proposed a quantum version of the famous Dijkstra's
shortest path algorithm, which was shown to give very
good performance with an appropriate selection of the routing
metric that considers the specific properties of quantum
networks~\cite{van_meter_path_2013}.
More recently, Caleffi~\textit{et al.} have proposed a slightly less efficient
variation of Dijkstra's algorithm that can work with non-isotonic routing
metrics, which they have advocated to provide superior performance in
selected use cases~\cite{caleffi_optimal_2017}.
Dijkstra's algorithm is also the subject of~\cite{chakraborty_distributed_2019},
where the authors lay some mathematical foundations that allow them to
derive upper bounds of performance in specific network topologies, including
grid and ring.

A different direction is explored by Pant~\textit{et al.}, who
studied the distribution of routing information to the
nodes~\cite{pant_routing_2019}; for this they propose a time-slotted
approach: in the first part of the slot every repeater tries to
create a local entanglement with all its neighbors, then in the
second part the paths are established as instructed by a centralized
authority.
One interesting aspect of the paper is that multiple paths are selected
for the same (source, destination) to maximize the rate of end-to-end Bell pairs.
We have also adopted this time-slotted model in \cite{cicconetti2021request},
where we have investigated the issue of ``scheduling'' of traffic flows, i.e.,
determining the order in which to assign paths to pending requests, in case
the network resources are not sufficient to serve them all.
This problem is called ``distribution'' in~\cite{dai_optimal_2020}, where
the authors formulate it as an \ac{ILP}, for which they derive closed
formula performance bounds in the case of a homogeneous chain of
quantum repeaters.
The issue is also addressed in~\cite{chakraborty_entanglement_2020},
where the authors have proposed to split the overall quantum routing
problem in two to reduce the computational complexity: first, they
determine the rates achievable by the traffic flows under the given
network constraints using an approach based on multi-commodity flow
optimization, then they map these rates to paths.
The paper adopts a network model using probabilistic entanglement swapping,
which we reuse in this work (described in \rsec{cont:model}).

An important reference for our study is~\cite{li_effective_2021}, where
the authors study the allocation strategy of traffic flows for which
the paths have been pre-determined: they do so by borrowing the fairness
concept from data networks and re-using traditional algorithms from
the relevant literature.
In our paper, we also borrow from the same literature, though we apply
the concepts to a different class of applications, as it will be clear
in the next section.
\textit{As a matter of fact, all the scientific works cited above
have focused on point-to-point traffic flows, while in \rsec{cont:traffic}
we introduce a different type of traffic that is more suitable to
model distributed \ac{QC}, with distinguishing features that do not
allow the reuse of state-of-the-art solutions.}
Rather, we claim that any existing routing/allocation/scheduling
solutions should work \textit{in parallel} to our proposed scheme
to provide an effective resource allocation to each of the two
traffic classes.

In addition to mere routing aspects, system-wide studies have also been
published.
We mention~\cite{van_meter_quantum_2021}, which is a compendium of
several previous studies from the same authors that illustrates an
overall architecture of the Quantum Internet, also including
application, protocol, and deployment aspects at a high level.
On the other hand, other works have focused on specific components,
which are complementary to the research activity presented, e.g.,
\cite{zhao_quantum_2021} on congestion control in transport protocols
and~\cite{dahlberg_link_2019} on the link layer, with a focus on
hardware and physical-layer considerations.
%

Furthermore, some research groups have been working to define the
basic principles of \textbf{distributed \ac{QC}}.
Parekh~\textit{et al.} have defined an elegant framework for the parallel
execution of a broad class of quantum algorithms on multiple
nodes~\cite{parekh_quantum_2021}, both using remote entanglement and
with \ac{LOCC} only, also studying in depth three classes of algorithms:
variational quantum eigensolver, low-depth quantum amplitude estimation,
and quantum k-means clustering.
In~\cite{cuomo_optimized_2021} the authors address the problem of
the efficient compilation of circuits for distributed \ac{QC}
by considering that some gate operations will be executed remotely, hence
with much different latency and reliability than on-chip operations.
The research of Dahlber~\textit{et al.} moved in the same direction
and went as far as defining a set of low-level instructions (called
NetQASM) for distributed \ac{QC} systems seamlessly
supporting local and remote gates~\cite{dahlberg_netqasm_2021}.
%
\textit{These works confirm that there is a growing interest in
distributed \ac{QC}, which is a motivation for our work.}
%

%

%
  \section{Traffic Classes in Quantum Networks and Resource Allocation}%
  \label{sec:contribution}%
  In this section we illustrate the novel contribution of our work.
We first introduce the system model used for quantum networks (\rsec{cont:model}).
Then, in \rsec{cont:traffic}, we elaborate on two different classes
of traffic, namely constant-rate flows and distributed \ac{QC}
applications.
For the latter we propose a resource allocation solution in
\rsec{cont:apps}.

\myssec{System model}{cont:model}

In this work we adopt the model proposed
in~\cite{chakraborty_entanglement_2020}, which focuses on some
fundamental characteristics of quantum networks, which will remain
true for at least the 1G of quantum repeaters (described in
\rsec{basics}).
We consider a network of quantum devices, called \textit{nodes},
interconnected via \textit{links}.
Without loss of generality, we do not distinguish between quantum
computers vs.\ repeaters.
The links are directional and they are assumed to have a fixed
capacity, in terms of the rate at which Bell pairs
can be generated, indicated as $C_{i,j}>0$, where $i$ and $j$ are the
two nodes connected.
For instance, $C_{1,2} = 8$ means that the quantum physical-layer devices
between node $1$ and $2$ are such that 8 maximally entangled Bell pairs
will be generated every second; we recall that qubits cannot be stored,
thus all unused Bell pairs will be necessarily discarded.
The network can then be represented as a weighted graph $\mathcal{G}$ where the
vertices $\mathcal{V}$ are the nodes, the edges $\mathcal{E}$ are the
links, and the weights are the link capacities.
In practice, depending on the technology used, the initial fidelity
of a Bell pair is a stochastic process with expectation
$F^\mathrm{init}_{ij}$, which could be used to decorate the edges
further; for simplicity of notation, in this work we assume that
$F^\mathrm{init}_{ij} = \bar{F}$ for all $(i,j) \in \mathcal{E}$.


As already explained, 1G quantum repeaters do not use quantum error
corrections, i.e., the measurement operation (see
\rfig{basics-3}), which enables entanglement swapping, may fail according
to some (as of yet) unknown probabilistic process: when this happens,
the qubits have to be discarded and end-to-end entanglement fails accordingly.
If linear optics are used to implement this operation, the failure probability
is \textit{at least} 50\%~\cite{sangouard_quantum_2011}.
%
Let us consider two nodes $i$ and $k$ attempting repeatedly end-to-end
entanglement with a rate $R$  through node $j$, which has $q_j$
average measurement success probability.
The entanglement will only succeed a fraction of the times ($q_j$), which
means that the rate of usable Bell pairs will be $r = R \cdot q_j$.
We call $r$ net rate, and $R$ gross rate.
The net rate is available to the quantum applications, while
the gross rate is that consumed on network resources, which has to be
available along the path in accordance with the link capacities:
$\frac{r}{q_j} \leq \min\{C_{ij},C_{jk}\}$.
If there were $N$ intermediate nodes, the net rate would decrease
exponentially: $r = R \cdot \prod_{j=1}^N q_j$, which is a well-known
undesirable property of quantum networks.
Again, to keep the notation simple, we assume that the average
measurement success probability of all nodes in the network is
equal, i.e., $q_j = q$ for all $j \in \mathcal{V}$, which gives:

\begin{equation}\label{eq:gross-rate}
    R = \frac{r}{q^L},
\end{equation}
where $L \geq 1$ is the length of the path $p$ between the two nodes
(in fact, with $L = 0$ there is no entanglement swapping and it is
$R = r$).
Also the fidelity is known to degrade exponentially with the path
size, according to the following formula~\cite{briegel_quantum_1998}
(under some simplifying assumptions):

\begin{equation}\label{eq:fidelity:perfect}
    F = \frac{1}{4} + \frac{3}{4} \left( \frac{4\bar{F}-1}{3} \right)^{L+1}.
\end{equation}

\myfigeps{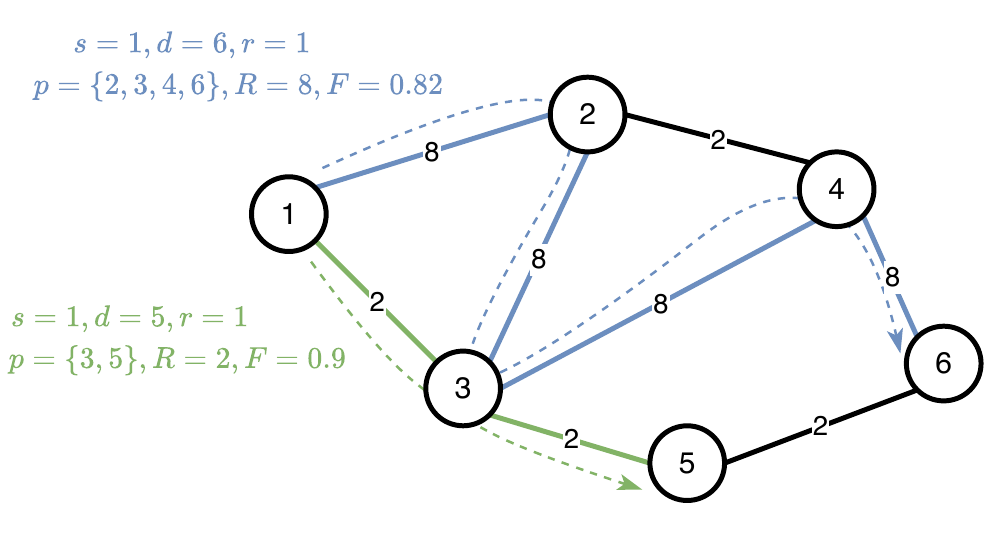}{%
Example quantum network, with $q = 0.5$ and $F = 0.95$.}

An example network is shown in \rfig{network-example}.
Assume we want to create a sequence of end-to-end entanglements between
nodes 1 and 6 with rate 1 Bell pair/s.
The shortest path, in number of hops, is through nodes 3 and 5, but
the path size would be $L = 2$, which entails $R = r / q^L = 1 / (0.5)^2 = 4$,
which exceeds the minimum capacity of the edges along the path, that is 2.
The only viable path is through 2, 3, and 4 (blue walk in the figure,
with $L = 3$), requiring a gross rate $R = 8$, which is available
on all the edges.
Note that going through this longer route consumes more network resources,
but it is inescapable in this case.
Instead, a path between nodes 1 and 5 with $r = 1$ \textit{is} feasible passing
through 3 (green walk in the figure), because with $L = 1$ the gross
rate is just 2, which is compatible with the capacities along the
edges (1,3) and (3,5).
%


\myssec{Quantum network traffic classes}{cont:traffic}

So far, the problem of routing demands in quantum networks has been
investigated only for a single class of applications, i.e., those
essentially defined by source, destination, and Bell rate (and possibly
minimum fidelity).
Typical applications that fit this model are: \ac{QKD}, i.e., the secure
exchange of a shared secret between two parties, and quantum sensing, for
the accurate measurement of physical quantities.
For those applications, both the rate and fidelity requirements are to
be intended as strict guarantees: if not matched, then the application
cannot work properly, but if they are exceeded then the performance does
not get any better.
Therefore, it is implicit that admission of new flows of this type must
be subject to an admission control scheme to verify that resources
are available in the network.

\begin{table}[tbp]%
\caption{%
Characteristics of constant-rate applications (flows)
vs. distributed \ac{QC} applications (apps).}%
\vspace{-2em}%
\centering%
\begin{tabularx}{\columnwidth}{*{3}{p{0.3\columnwidth}}}
    \hline

    \thead{} &
    \thead{Flows} &
    \thead{Apps} \\

    \hline

    \textit{Nodes involved} &
    source + destination &
    host + set of peers \\

    \textit{Rate requirement} &
    constant Bell rate &
    elastic Bell rate \\

    \textit{Fidelity requirement} &
    strict &
    flexible \\

    \textit{Session duration} &
    user-driven &
    depends on allocation \\

    \textit{QoS} &
    guaranteed upon admission &
    best-effort \\


    \hline
\end{tabularx}%
\label{tab:traffic}%
\vspace{-1em}%
\end{table}%

Distributed \ac{QC} is another class of applications, which to the best of
our knowledge has been totally neglected so far, from the point of view
of quantum routing.
Applications in this class do not have a minimum required rate, but rather they
can greedily consume as many Bell pairs as provided, under the assumption
that end-to-end entanglement is the limiting factor for the execution
of quantum circuits on distributed infrastructures, which appears to
be reasonable in the near future based on current technology trends.
Furthermore, a single host wishing to offload its computation to other
\acp{QC} may want to do so in a pool of peers, rather than point-to-point:
this adds a degree of freedom in the routing decision as one does not only
have to select a path, but also choose which of the possible destinations
to activate for a given application.
In the remainder of this paper, we call for short \textit{flows} the
constant-rate traffic flows vs.\ \textit{apps} the flexible distributed
computing applications.
A comparison of their main distinguishing features is reported in
\rtab{traffic}.

\myfigeps{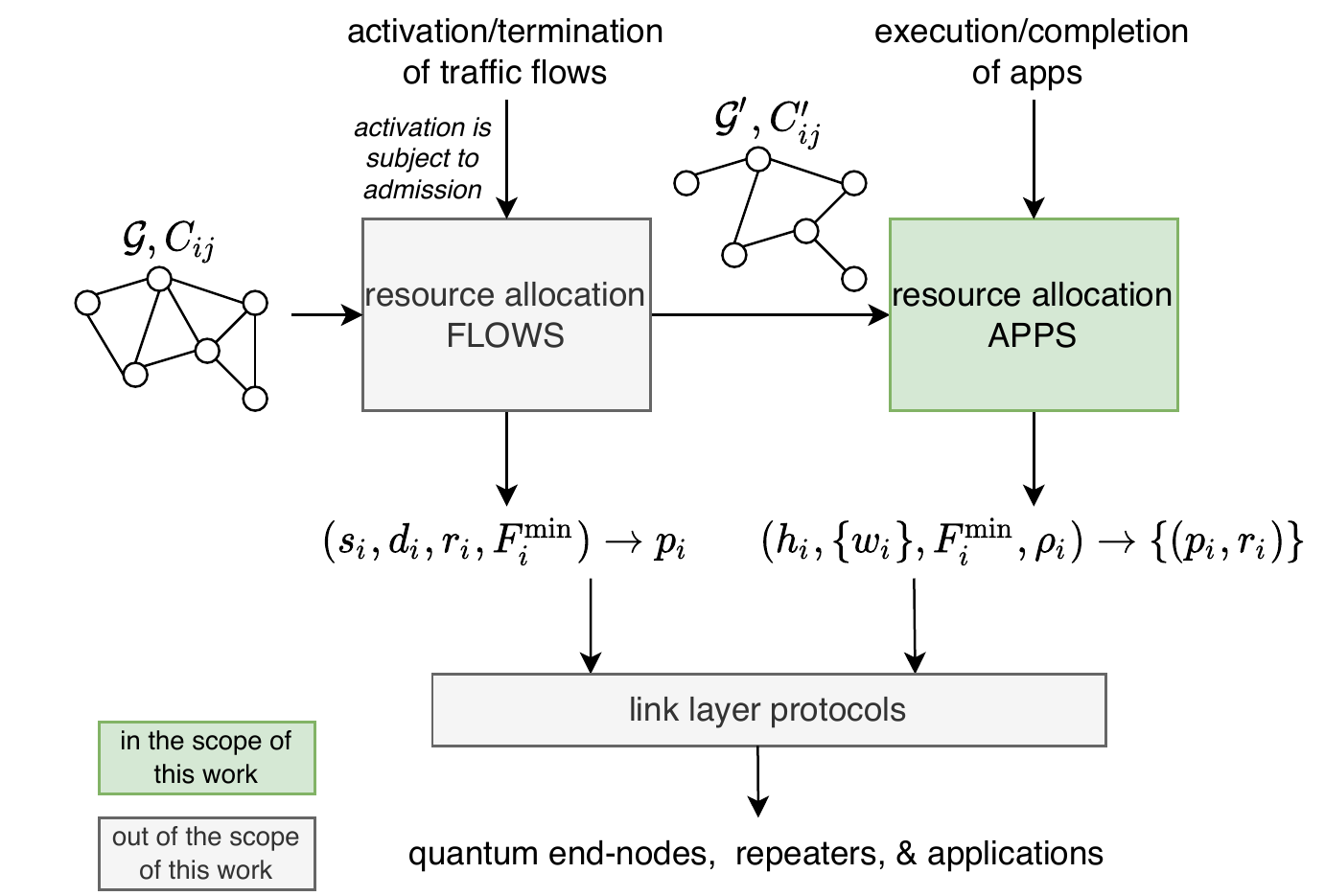}{%
Overview of the architectural elements for the coexistence of
the two traffic classes considered in this work for quantum networks.}

We can speculate that in a real quantum network flows and apps will coexist,
much like latency-sensitive constant bit-rate applications (e.g., digitized
voice) coexist with best-effort elastic TCP applications in the Internet today.
Only time will tell whether one type of traffic will dominate or which
business model will emerge, but for now we argue that it makes sense to
give priority to flows: since their requirements are inflexible, the rate
of flows that fail admission can be minimized by considering all
the network resources available for this class of traffic.
Accordingly, we propose a high-level view of the overall resource allocation,
illustrated in \rfig{system-1}.
In the figure the difference between flows and apps is once more evident:
a flow is identified by $(s,d,r,F^{\min})$, i.e., source, destination,
rate, and minimum fidelity, and it is mapped to a single path $p$;
an app is identified by $(h,\{w\},F^{\min},\rho)$, i.e., host, set of
candidate peers (depending on the commercial agreements between
\ac{QC} infrastructures), minimum fidelity, and priority, and it is mapped
to a set of paths, each with a given net rate.
We have identified the priority (as a numerical weight) as a means
to provide differentiated service; for example, a higher priority
may be acquired at extra service costs or it can be an internal
parameter that increases while the resources are not used.
The green box in the figure, i.e., the resource allocation of apps,
is elaborated further in the next section, while the gray boxes are
covered already by prior works:
\textit{resource allocation of flows} is investigated in 
\cite{van_meter_path_2013,caleffi_optimal_2017,chakraborty_distributed_2019,cicconetti2021request,dai_optimal_2020,chakraborty_entanglement_2020,li_effective_2021}, whereas 
\textit{link layer protocols} are explored in
\cite{pant_routing_2019,dahlberg_link_2019,zhao_quantum_2021}.

\myssec{Resource allocation of apps}{cont:apps}

For apps, the resource allocation problem can be expressed in general
terms as follows: given a set of apps, find the set of paths, and
corresponding rate, for each possible (host, peer), that maximizes
a given objective function.
In this form, it resembles the classical problem of resource allocation
in data networks (e.g., \cite{bertsekas_data_1992}), even though there
are important new features: i) the gross capacity decreases as the
path length increases (more resources needed) and ii) so does the fidelity
(which may eventually result in the path becoming unfeasible).
Nonetheless, we believe that the notion of \textit{fairness} widely
studied in data networks can be swiftly reused in this context as
the resource allocation goal.
To this end, we propose an algorithm for the resource allocation
of apps inspired from well-known \ac{WRR}~\cite{105173}.
\textit{The basic idea is to pre-compute, for each app, all the
possible paths towards the requested destinations (up to a maximum
of $k$ paths per destination, to keep the computational complexity
low), then to allocate resources to apps in round-robin, for fairness
reasons, where the maximum amount of gross rate an app is given in
each round is a fraction of the round size $\phi$ proportional
to its weight.}
The algorithm is explained with the help of the pseudo-code in \ralgo{wrr}.
Full implementation details can be found in the online repository of the
simulator source code, publicly available as described in \rsec{eval}.

\begin{algorithm}
    {\small
    \DontPrintSemicolon
    $\mathcal{G}'$ is the residual graph after all the possible flows have been allocated\;
    
    $\mathcal{A}$ is the set of apps to be provisioned, with each app $i$ characterized by $(h_i,\mathcal{W}_i,F^{\min}_i,\rho_i)$\;

    $\forall i \in \mathcal{A}: \mathcal{P}_i \leftarrow \emptyset$\;

    $\mathcal{L} \leftarrow \{ i | i \in \mathcal{A} \}$\;

    $a \leftarrow \texttt{last}(\mathcal{L})$\;

    $\delta \leftarrow 0$\;
    
    $\mathcal{F} \leftarrow \emptyset$\;

    \For{$i \in \mathcal{A}$}{
        \For{$j \in \mathcal{W}_i$}{
            $\mathcal{P}_i \leftarrow \mathcal{P}_i \cup \texttt{find\_paths}(h_i, w_{ij}, k, \mathcal{G}')$\;
        }
    }

    \While{$\mathcal{L} \neq \emptyset$}{
        \If{$\delta = 0$}{
            $a \leftarrow \texttt{next}(a, \mathcal{L})$\;
            $\delta \leftarrow \phi \frac{\rho_a}{\sum_{i \in \mathcal{A}} \rho_i}$\;
        }
        $p \leftarrow \emptyset$\;
        \While{$p = \emptyset \wedge \mathcal{P}_a \neq \emptyset$}{
            $p \leftarrow \texttt{find\_shortest\_path}(\mathcal{P}_a)$\;
            \If{$\exists e \in p | e \notin \mathcal{G}'$}{
                $\texttt{remove\_path}(p, \mathcal{P}_a)$\;
                $p \leftarrow \emptyset$\;
            }
        }
        \eIf{$p = \emptyset$}{
            $a \leftarrow \texttt{remove\_app}(a, \mathcal{L})$\;
            $\delta \leftarrow \phi \frac{\rho_a}{\sum_{i \in \mathcal{A}} \rho_i}$\;
        }{
            $R = \min\left\{\delta, \min_{e \in p} C_e\right\}$\;
            $\texttt{add}(a, p, R, \mathcal{F})$\;
            $\texttt{remove\_capacity}(R, p, C_e)$\;
            $\texttt{update\_graph}(\mathcal{G}', C_e)$\;
            $\delta \leftarrow \delta - R$\;
        }
    }

    \KwRet{$\mathcal{F}$}
    }

    \caption{Resource allocation of apps.}
    \label{algo:wrr}
\end{algorithm}

The algorithm takes as input the residual graph $\mathcal{G}'$ after
all the possible flows have been allocated and a set $\mathcal{A}$
of apps each characterized by $(h_i,\mathcal{W}_i,F^{\min}_i,\rho_i)$,
with $i \in \mathcal{A}$; it produces as output the set of paths
$\mathcal{F}$ that are allocated to the apps, where any app may
have multiple paths each with a different net/gross rate.
The algorithm has two system parameters: $k \in \mathbb{N}$ and
$\phi$, in Bell pairs/s, which we call \textit{round size}, already introduced.
The following working variables are used:
\begin{myitemlist}
    \item $\mathcal{L}$ is the
    \textit{active list}, i.e., the set of apps that can be still
    provisioned: it is initialized with all the input apps (line~4) and
    then slowly depleted (line~27) each time an app $a$ does not have
    further paths available in the residual graph;
    \item $a$ is the current
    app considered for allocation, initialized with the last app in the
    active list (line~5);
    \item $\delta$, also called \textit{credit} below, is the maximum
    amount of gross capacity that can be allocated to the current
    app $a$;
    \item $\mathcal{P}_a$
    is the set of paths available for the current app $a$, which is
    initialized at the beginning of the algorithm (lines~8--12) and
    then depleted when one path $p$ is not feasible anymore in the
    residual graph (line~22).
\end{myitemlist}

\begin{table}[tbp]%
\caption{Functions used in the pseudo-code of \ralgo{wrr}.}%
\vspace{-2em}%
\centering%
\begin{tabularx}{\columnwidth}{p{0.46\columnwidth}|p{0.47\columnwidth}}
    \hline

    \texttt{last($\mathcal{L}$)} &
    return the last app in the active list \\
 
    \texttt{next($a$,$\mathcal{L}$)} &
    return the subsequent app to $a$
    in the active list $\mathcal{L}$, wrapping around when the end is reached \\

    \texttt{remove\_app($a$,$\mathcal{L}$)} &
    remove the app $a$ from $\mathcal{L}$ and return the next app in the list,
    wrapping around when the end is reached \\

    \texttt{find\_paths($s$,$d$,$k$,$\mathcal{G}$)} &
    returns the set of (up to) $k$ paths from node $s$ to node $d$
    in $\mathcal{G}$ \\

    \texttt{find\_shortest\_path($\mathcal{P}$}) &
    return the shortest
    path in $\mathcal{P}$ \\

    \texttt{remove\_path($p$,$\mathcal{P}$)} &
    remove the path $p$ from the
    set $\mathcal{P}$ \\

    \texttt{add($a$,$p$,$R$,$\mathcal{F}$)} &
    assign the path $p$ to app $a$ with
    gross rate $R$ in the output $\mathcal{F}$; if the same path $p$ is already
    present in $\mathcal{F}$ for app $a$, then $R$ is simply added to the
    gross rate already reserved \\

    \texttt{remove\_capacity($R$, $p$,$\mathcal{C}_e$)} &
    remove the capacity $R$ from
    the capacities $\mathcal{C}_e$ along the path $p$ \\

    \texttt{update\_graph($\mathcal{G}$,$\mathcal{C}_e$)} &
    update $\mathcal{G}$ by removing the edges with vanishing capacity in $\mathcal{C}_e$ \\

    \hline
\end{tabularx}%
\label{tab:functions}%
\vspace{-1em}%
\end{table}%

\noindent\ul{Pseudo-code walk through.}
Lines~1--12 cover the \textbf{input definition and initialization} of the
working variables, as explained above.
The rest of the procedure is a loop that iterates over the active list
until it is empty (condition line~13.)
Within the main loop (lines~14--35), $a$ is the identifier of the current
app, which will be either assigned a path or removed from the active list.
The \textbf{first step} (lines~14--17) is to move to the next app $a$ if the credit $\delta$
of the current one is exhausted, in which case the credit of the new app
is initialized as $\delta = \phi \frac{\rho_a}{\sum_{i
\in \mathcal{A}} \rho_i}$, i.e., a fraction of the round size $\phi$
proportional to its priority.
The \textbf{second step} (lines~18--25) is to identify one possible path to
be assigned to the current app $a$; this is done by selecting the
shortest path $p$ from $\mathcal{P}_a$, which contains the list of
possible paths from $h_a$ to any of its possible destinations in
$\mathcal{W}_a$.
The shortest path $p$ is the one that requires the least amount of resources
among those available for the current app, according to \req{gross-rate},
\textit{and} gives the maximum fidelity, according to \req{fidelity:perfect}.
Since $\mathcal{P}_a$ was determined with the initial graph $\mathcal{G}'$, and
the latter will be updated as new paths are assigned to the apps, it is
possible that $p$ is no longer feasible under the updated graph: in this case,
the path is removed from $\mathcal{P}_a$ (line~22) and the loop is restarted.
When the loop is done (line~25), the path $p$ is either valid or there are not
anymore feasible paths for the current app $a$.
The two cases are handled in the \textbf{third and last step} (lines~26--35).
If the app $a$ does not have any more feasible paths (line~26), it is removed from the
active list (line~27) and the credit for the next app is updated (line~28).
Otherwise, if the path $p$ is valid, the gross rate $R$ is first determined
as the minimum of the residual credit $\delta$ and the minimum capacity along
the path (line~30).
Then, the path is added to the output (line~31), the selected gross rate $R$ is
removed from the capacities of all the edges along the path (line~32), the
residual graph $\mathcal{G}'$ is updated to reflect the new capacities (line~33),
and the credit is updated (line~34).
The functions called in \ralgo{wrr} are defined in \rtab{functions}.


%
We now analyze the worst-case time complexity of \ralgo{wrr}, separately
for the initialization (lines~8--12) and the main loop (lines~13--35).
With regard to the \textbf{initialization}, if $A$ is the number of apps and
$W$ the average number of peers per host, we execute $A \cdot W$ times
the algorithm \texttt{find\_paths}.
The latter can be implemented with Yen's algorithm to determine the
$k$ shortest paths in a graph~\cite{yen_finding_1971}, which with
the help of a Fibonacci heap to keep edges
sorted~\cite{fredman_fibonacci_1984} is known to have complexity
$\mathcal{O}\left(k V (E+V \log V)\right)$.
%
%
%
A trade-off exists: if a large value of $k$ is used then more paths will
be found and the algorithm will be slower, but in the main loop there will
be a wider choice of paths to be considered for each host-destination pair.
Regarding the \textbf{main loop}, an exact upper bound of the time complexity is more
difficult to derive, because it depends on the specific network topology
and capacities.
In particular, the number of iterations of lines~13--36 may depend on either
the smallest capacity of edges along the candidate path or the credit given
to each app, which in turn depends on the value of the system parameter $\phi$.
Therefore, we can expect that another trade-off exists: smaller values of
$\phi$ lead to a slower algorithm, but also to
a better fairness.
This will be confirmed by the simulation results in the next section.
%

  \section{Performance Evaluation}%
  \label{sec:eval}%

In this section we evaluate the performance of the resource allocation
solution in \rsec{cont:apps} via simulation.
In the research community some simulation tools have been used and made
publicly available to evaluate the performance of quantum networks.
For example
NetSquid\footnote{\url{https://netsquid.org/}}~\cite{coopmans_netsquid_2021}
and
SeQUeNCe\footnote{\url{https://github.com/sequence-toolbox/SeQUeNCe}}~\cite{wu_sequence_2021}
are Python event-driven simulators that allow the user to customize
the basic building blocks of quantum networks for the evaluation
of protocols (e.g., quantum routing) and distributed applications
(e.g., \ac{QKD});
QuISP\footnote{\url{https://github.com/sfc-aqua/quisp}}~\cite{satoh_quisp_2021},
instead, is a full-fledged network simulator based on
Omnet++\footnote{\url{https://omnetpp.org/}} designed to study the
dynamic behavior of large-scale quantum networks, e.g., congestion
and stability aspects.
However, to the best of our knowledge, none of the public simulators
available so far address network resource provisioning, which is
the subject of our work.
Therefore, we have implemented a custom simulator, developed in C++
and using the Boost Graph Library~\cite{siek_boost_nodate}, which
is available as open source under a MIT license on
GitHub\footnote{\url{https://github.com/ccicconetti/quantum-routing/}, tag
\texttt{v1.0}.}; for full reproducibility, we have also included
the scripts to run the experiments, as well as the artifacts obtained
and the Gnuplot files to produce the plots.

Like in~\cite{dai_optimal_2020}, we use a \ac{PPP} to generate the
position of an average of $\mu$ nodes in a flat square grid with
edge size 60~km; a link is added between two nodes with probability
$p_{\mathrm{link}}$ if their Euclidean distance is smaller than a
threshold $\tau$.
The capacity of each link is drawn from a r.v. uniformly distributed
between 1~Bell pair/s and 400~Bell pairs/s, as
in~\cite{chakraborty_entanglement_2020}.
The initial fidelity of Bell pairs is $\bar{F} = 0.95$.
Based on calibration experiments (not shown but available in the
repository) we have selected $p_\mathrm{link}=0.5$ and two possible
values for $\mu$ (50 and 100) and $\tau$ (15~km and 20~km), which
yield networks with very different characteristics reported in \rtab{graphs}.

\begin{table}[tbp]%
\caption{Characteristics of the scenarios (average values).}%
\vspace{-2em}%
\centering%
\begin{tabularx}{\columnwidth}{*{5}{p{0.15\columnwidth}}}
    \hline

    &
    $\mu=100,$ $\tau=15$~km &
    $\mu=100,$ $\tau=20$~km &
    $\mu=50,$ $\tau=15$~km &
    $\mu=50,$ $\tau=20$~km \\

    \hline

    \textit{Capacity} &
    159k &
    257k &
    47k &
    67k \\

    \textit{Num edges} &
    792 &
    1282 &
    232 &
    331 \\

    \textit{Max degree} & 
    16 &
    24 &
    9 &
    13 \\

    \textit{Diameter} &
    8 &
    5 &
    10 &
    6 \\

    \hline
\end{tabularx}%
\label{tab:graphs}%
\vspace{-1em}%
\end{table}%

\begin{table}[tbp]%
\caption{Metrics used for the evaluation.}%
\vspace{-2em}%
\centering%
\begin{tabularx}{\columnwidth}{lX}
    \hline

    \thead{Metric} &
    \thead{Definition} \\

    \hline

    \textit{net rate} &
    sum of the net rates assigned to peers, in Bell pairs/s \\

    \textit{fidelity} &
    \req{fidelity:perfect} weighted on the net rates assigned to peers \\

    \textit{visits} &
    loop counter of Step~1-Step~6 in \rsec{cont:apps} \\

    \textit{fairness} &
    Raj Jain's index on net rates $\left(\bar{r}\right)^2/\left(\bar{r^2}\right)$ \\

    \hline
\end{tabularx}%
\label{tab:metrics}%
\vspace{-1em}%
\end{table}%

We used a Monte Carlo approach: for any combination of the parameters
under study, we simulated 10,000 drops with randomly generated networks and random workload.
The latter is made of a number of apps that depends on the experiment,
with host node ($h$) selected uniformly from all the nodes, and a set of peers
of cardinality $W$ sampled from all the nodes that can be reached by $h$
in at most $D$ hops.
We set $F^{\min} = 0$ for all apps, so that fidelity does not
constrain resource allocation and can be evaluated \textit{a
posteriori}.
Finally, it is always $\rho = 1$.
The impact of constrained fidelity and heterogeneous priorities is left
for future work.
The metrics used are reported in \rtab{metrics}.
%
Statistical significance has been verified for all the metrics in
the experiments performed, but we do not include error bars in plots
for better readability.

\myfigdoubleeps{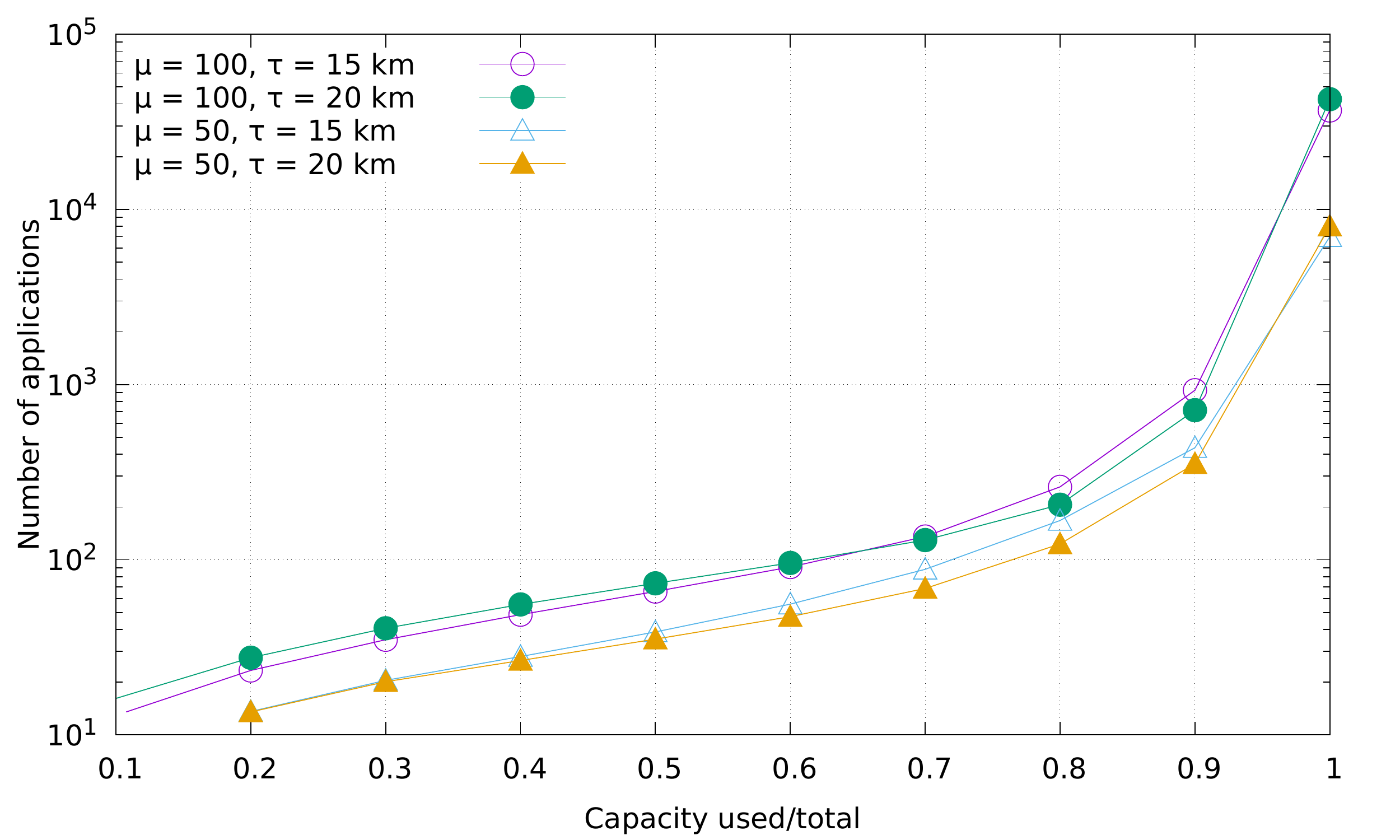}{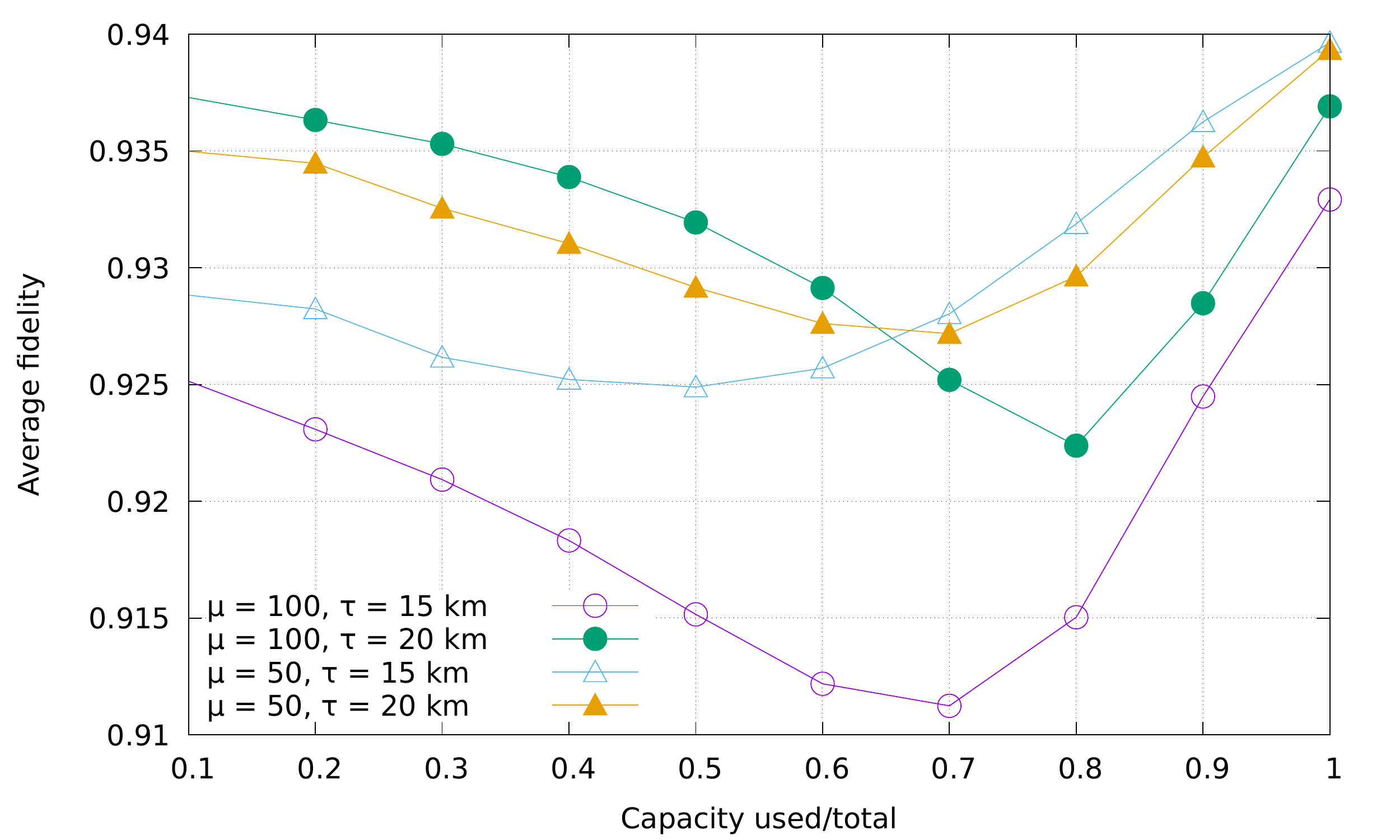}{experiment-002}{%
Experiment with increasing apps until a target
residual capacity is met, with
$W = 10$, $D = 5$, $k = 4$, $\phi = 10$~Bell pairs/s.}

We start with a first experiment where we increasingly add new apps
until we meet a target residual capacity, defined as the sum of the $C_{ij}$
of all the links at the end of the resource allocation.
In \rfig{experiment-002} (left) we show the number of apps reached vs.\ the
normalized residual rate from 0.1 (= only 90\% of the capacity remains unused
at the end of the allocation) to 1 (= fully used capacity).
As can be seen, all the curves increase very steeply (the y-axis is in
logarithmic scale) for low residual capacity: this shows that, despite the
elastic allocation of resources to apps, it is very difficult to exhaust
resources in a quantum network.
This is because \textit{reaching the capacity of a link through multiple hops 
requires an exponential increase of the resources used}, as in
\req{gross-rate}, which is in stark contrast with classical data networks.
In the right part of \rfig{experiment-002} we report the fidelity.
Interestingly, all the network combinations lead to a non-monotone
pattern: when there is abundance of resources (low x-axis values)
then only the best (i.e., shortest) paths are used; at intermediate
loads, resource allocation tends to explore also longer paths, as
$F^{\min} = 0$ in the experiments; finally, with many applications,
the only viable solution to reach all capacity is via short paths,
which on average increases the fidelity again.

\begin{table}[tbp]%
\caption{%
Qualitative results with 100 apps varying $W,D,k,\phi$.}%
\vspace{-2em}%
\centering%
\begin{tabularx}{0.8\columnwidth}{*{5}{c}}
    \hline

    Metric &
    $W$ &
    $D$ &
    $k$ &
    $\phi$ \\

    \hline

    \textit{net rate} &
    $\uparrow$ linear &
    $\downarrow$ asynt &
    $\downarrow$ slight &
    $\approx$  \\

    \textit{fidelity} &
    $\uparrow$ steep &
    $\downarrow$ asynt &
    $\approx$ &
    $\approx$ \\

    \textit{visits} &
    $\uparrow$ linear &
    $\downarrow$ asynt &
    $\uparrow$ linear &
    $\downarrow$ exp \\

    \textit{fairness} &
    $\uparrow$ steep &
    $\downarrow$ asynt &
    $\approx$ &
    $\approx$ then $\downarrow$ exp \\

    \hline
\end{tabularx}%
\label{tab:002-sensitivity}%
\vspace{-1em}%
\end{table}%

\myfigdoubleeps{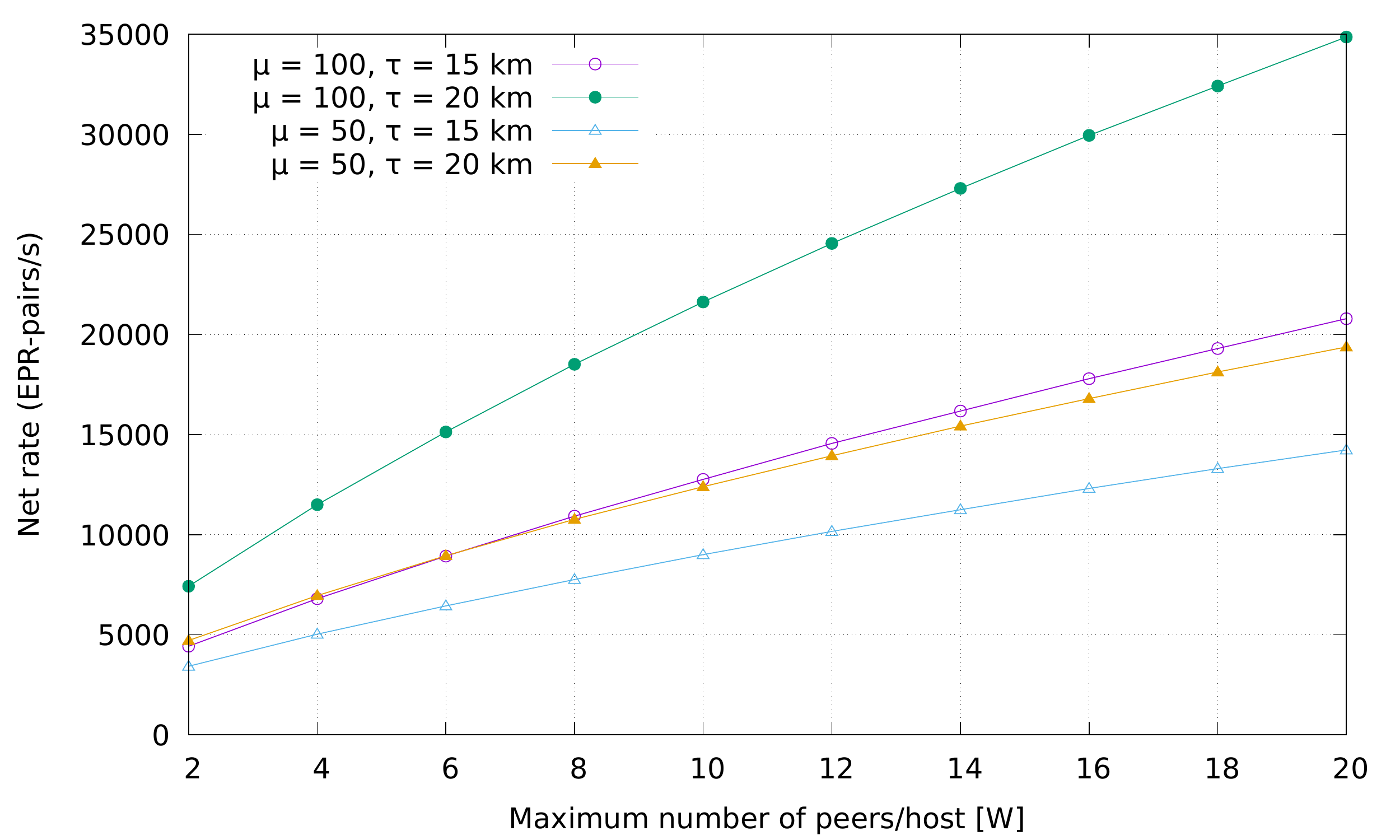}{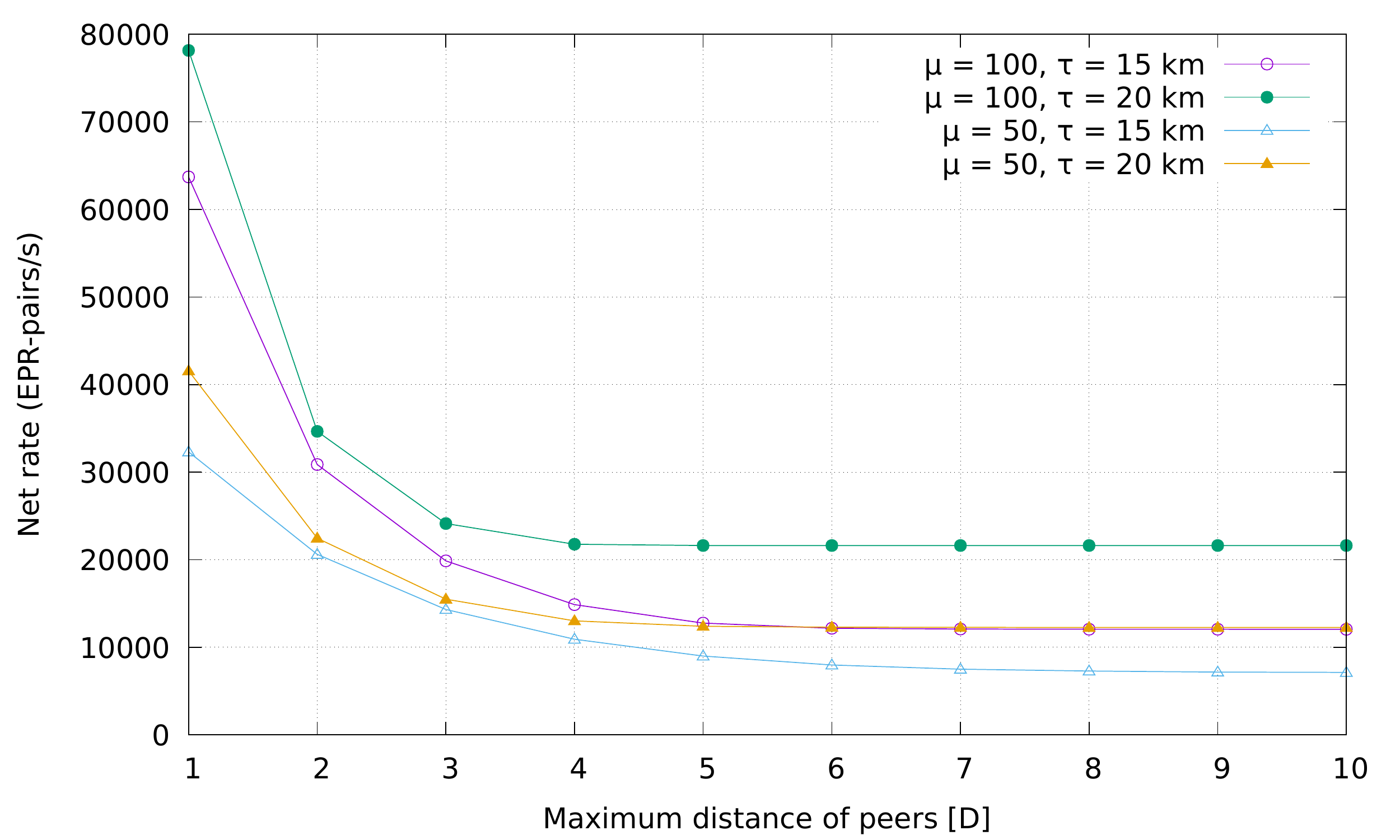}{002-sens-netrate}{%
Net rate when increasing $W$ (left) and $D$ (right).}

\myfigdoubleeps{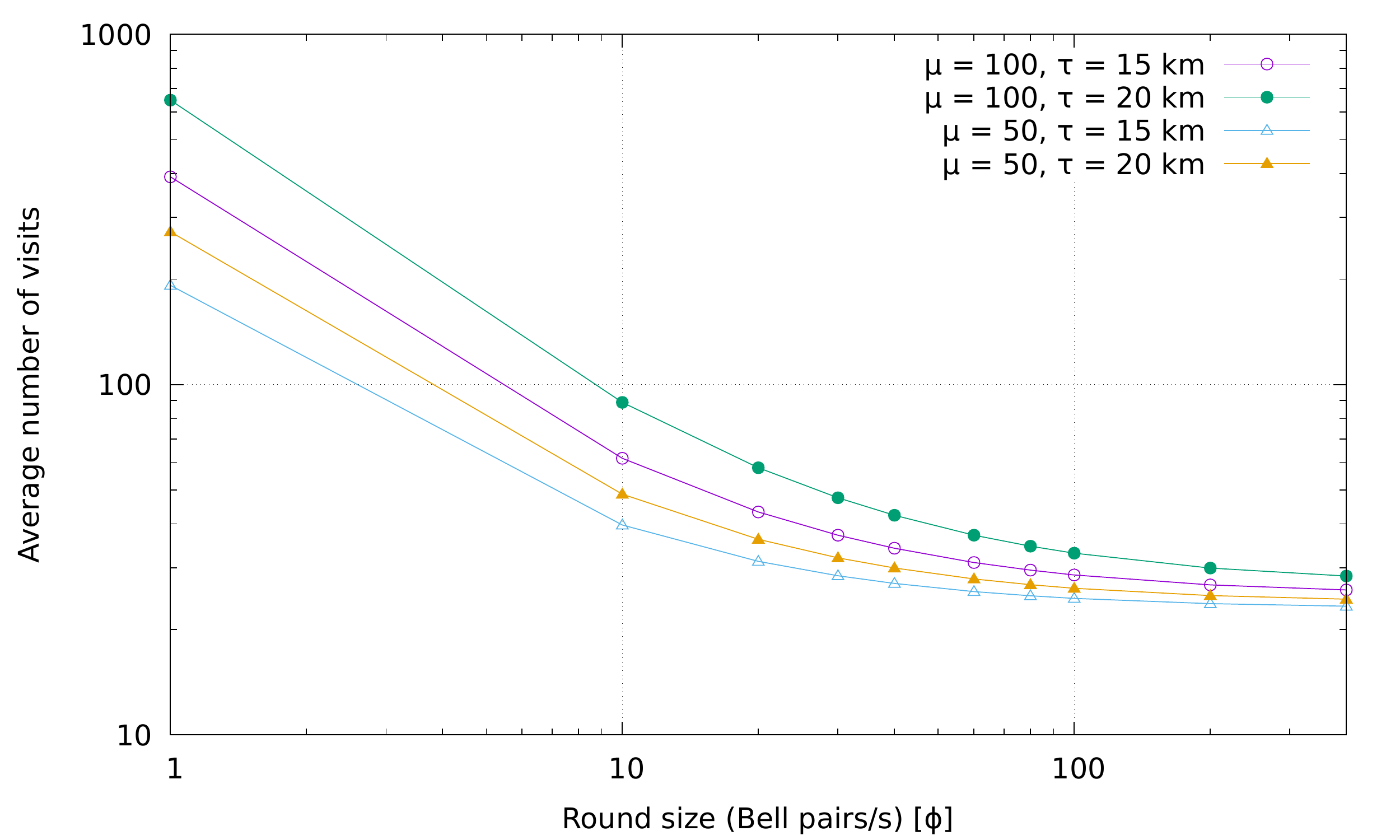}{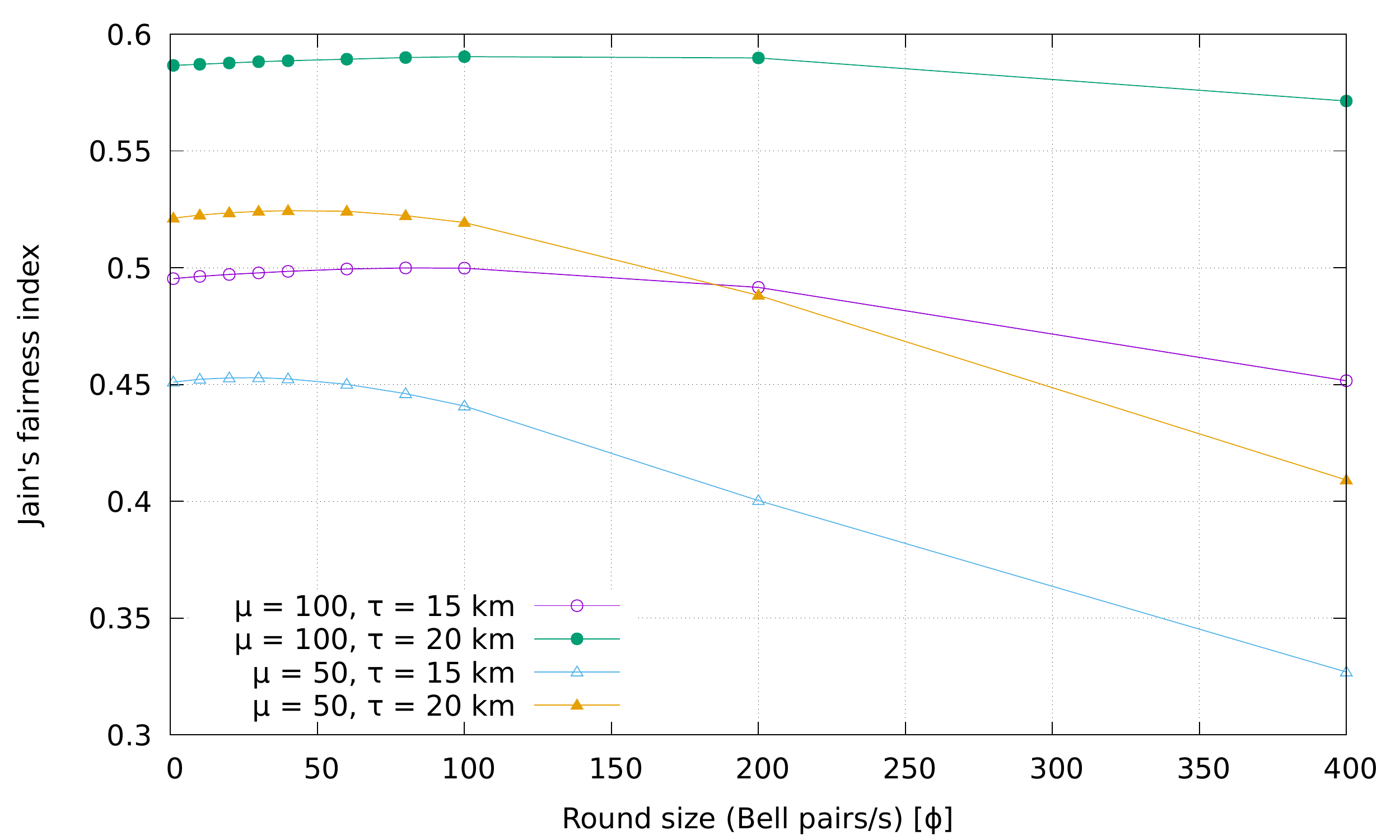}{002-sens-q}{%
Number of visits (left) and fairness Jain's index (right) when increasing $\phi$.}

In a second batch of experiments we study the performance trade-offs
of four key parameters by varying each one in a range while keeping the
others constant to keep the analysis tractable.
In particular, we simulate $W \in \{2,\ldots,20\}$,
$D \in \{1,\ldots,10\}$,
$k \in \{2,\ldots,20\}$,
and $\phi/100 \in \{1,\ldots,400\}$~Bell pairs/s;
the constant values are: $W=10$, $D=5$, $k=4$, and $\phi/100 = 10$~Bell pairs/s.
We report in \rtab{002-sensitivity} how the increase of each key parameter
affects the performance, from which we gather the following remarks:

\begin{myitemlist}
    \item Having a larger pool of peers (i.e., large $W$) is generally
    beneficial in terms of all the user-oriented performance metrics,
    but the resource allocation time complexity increases, so if
    execution time is a bottleneck then a trade-off arises.
    The net rate is shown in the left part of \rfig{002-sens-netrate},
    which also indicates a clear correlation with the network density.

    \item The peers should be as close as possible to their respective hosts.
    In particular, as can be seen in \rfig{002-sens-netrate} (right part),
    adding peers with distance above 4-5 does not yield noticeable
    improvements in terms of the net rate, for all the combinations of
    the network parameters.
    We note that in practice an app's peers may
    depend on external factors, such as contracts between \ac{QC}
    infrastructures and interface/technology compatibility limitation,
    so this parameter cannot be considered fully under the control
    of the quantum network operator or the users.

    \item $k$ can be small, as this yields both a higher net rate \textit{and}
    reduces time complexity.

    \item As expected, a trade-off exists between the time complexity and
    fairness in the choice of $\phi$, which is a purely internal parameter
    of the algorithm.
    This is studied in quantitative terms in \rfig{002-sens-q},
    suggesting that intermediate values achieve an excellent compromise.
\end{myitemlist}

%

%
  \section{Conclusions}%
  \label{sec:conclusions}%
  In this work we have first provided an introduction to
the emerging topics of quantum computing and networking.
Then, we have raised awareness on distributed \ac{QC}, which is 
a class of quantum applications that has not received significant attention
so far compared to point-to-point applications, like \ac{QKD} and sensing.
We have elaborated on the distinguishing features of the two classes
and proposed a resource allocation scheme for distributed \ac{QC}
that takes into account the fundamental aspects of end-to-end
entanglement in quantum networks.
Its performance has been evaluated with
simulations, which have 
led to the identification of critical trade-offs, which will have to
be studied in future research for efficient deployment and run-time
optimization of quantum networks.
Further open research areas are: the use of purification
to increase fidelity at the expense of capacity; modeling distributed
\ac{QC} applications to understand their characteristic time scales
and requirements; integration with link layer protocols.%

\section*{Acknowledgment}

Work co-funded by EU, \textit{PON Ricerca e Innovazione} 2014--2020
FESR/FSC Project ARS01\_00734 QUANCOM, and European High-Performance
Computing Joint Undertaking (JU) under grant agreement No 101018180
HPCQS.




\begin{acronym}
  \acro{3GPP}{Third Generation Partnership Project}
  \acro{5G-PPP}{5G Public Private Partnership}
  \acro{AA}{Authentication and Authorization}
  \acro{ADF}{Azure Durable Function}
  \acro{AI}{Artificial Intelligence}
  \acro{API}{Application Programming Interface}
  \acro{AP}{Access Point}
  \acro{AR}{Augmented Reality}
  \acro{BGP}{Border Gateway Protocol}
  \acro{BSP}{Bulk Synchronous Parallel}
  \acro{BS}{Base Station}
  \acro{CDF}{Cumulative Distribution Function}
  \acro{CFS}{Customer Facing Service}
  \acro{CPU}{Central Processing Unit}
  \acro{DAG}{Directed Acyclic Graph}
  \acro{DHT}{Distributed Hash Table}
  \acro{DNS}{Domain Name System}
  \acro{ETSI}{European Telecommunications Standards Institute}
  \acro{FCFS}{First Come First Serve}
  \acro{FSM}{Finite State Machine}
  \acro{FaaS}{Function as a Service}
  \acro{GPU}{Graphics Processing Unit}
  \acro{HTML}{HyperText Markup Language}
  \acro{HTTP}{Hyper-Text Transfer Protocol}
  \acro{ICN}{Information-Centric Networking}
  \acro{IETF}{Internet Engineering Task Force}
  \acro{IIoT}{Industrial Internet of Things}
  \acro{ILP}{Integer Linear Programming}
  \acro{IPP}{Interrupted Poisson Process}
  \acro{IP}{Internet Protocol}
  \acro{ISG}{Industry Specification Group}
  \acro{ITS}{Intelligent Transportation System}
  \acro{ITU}{International Telecommunication Union}
  \acro{IT}{Information Technology}
  \acro{IaaS}{Infrastructure as a Service}
  \acro{IoT}{Internet of Things}
  \acro{JSON}{JavaScript Object Notation}
  \acro{K8s}{Kubernetes}
  \acro{KVS}{Key-Value Store}
  \acro{LCM}{Life Cycle Management}
  \acro{LL}{Link Layer}
  \acro{LOCC}{Local Operations and Classical Communication}
  \acro{LTE}{Long Term Evolution}
  \acro{MAC}{Medium Access Layer}
  \acro{MBWA}{Mobile Broadband Wireless Access}
  \acro{MCC}{Mobile Cloud Computing}
  \acro{MEC}{Multi-access Edge Computing}
  \acro{MEH}{Mobile Edge Host}
  \acro{MEPM}{Mobile Edge Platform Manager}
  \acro{MEP}{Mobile Edge Platform}
  \acro{ME}{Mobile Edge}
  \acro{ML}{Machine Learning}
  \acro{MNO}{Mobile Network Operator}
  \acro{NAT}{Network Address Translation}
  \acro{NISQ}{Noisy Intermediate-Scale Quantum}
  \acro{NFV}{Network Function Virtualization}
  \acro{NFaaS}{Named Function as a Service}
  \acro{OSPF}{Open Shortest Path First}
  \acro{OSS}{Operations Support System}
  \acro{OS}{Operating System}
  \acro{OWC}{OpenWhisk Controller}
  \acro{PMF}{Probability Mass Function}
  \acro{PPP}{Poisson Point Process}
  \acro{PU}{Processing Unit}
  \acro{PaaS}{Platform as a Service}
  \acro{PoA}{Point of Attachment}
  \acro{PPP}{Poisson Point Process}
  \acro{QC}{Quantum Computing}
  \acro{QKD}{Quantum Key Distribution}
  \acro{QoE}{Quality of Experience}
  \acro{QoS}{Quality of Service}
  \acro{RPC}{Remote Procedure Call}
  \acro{RR}{Round Robin}
  \acro{RSU}{Road Side Unit}
  \acro{SBC}{Single-Board Computer}
  \acro{SDK}{Software Development Kit}
  \acro{SDN}{Software Defined Networking}
  \acro{SJF}{Shortest Job First}
  \acro{SLA}{Service Level Agreement}
  \acro{SMP}{Symmetric Multiprocessing}
  \acro{SoC}{System on Chip}
  \acro{SLA}{Service Level Agreement}
  \acro{SRPT}{Shortest Remaining Processing Time}
  \acro{SPT}{Shortest Processing Time}
  \acro{STL}{Standard Template Library}
  \acro{SaaS}{Software as a Service}
  \acro{TCP}{Transmission Control Protocol}
  \acro{TSN}{Time-Sensitive Networking}
  \acro{UDP}{User Datagram Protocol}
  \acro{UE}{User Equipment}
  \acro{URI}{Uniform Resource Identifier}
  \acro{URL}{Uniform Resource Locator}
  \acro{UT}{User Terminal}
  \acro{VANET}{Vehicular Ad-hoc Network}
  \acro{VIM}{Virtual Infrastructure Manager}
  \acro{VR}{Virtual Reality}
  \acro{VM}{Virtual Machine}
  \acro{VNF}{Virtual Network Function}
  \acro{WLAN}{Wireless Local Area Network}
  \acro{WMN}{Wireless Mesh Network}
  \acro{WRR}{Weighted Round Robin}
  \acro{YAML}{YAML Ain't Markup Language}
\end{acronym}

\end{document}